\documentclass[useAMS,usenatbib,12pt,epsfig]{mn2e}
\usepackage[english]{babel}
\usepackage{amssymb}
\usepackage{graphicx}

\title[HI content and dust-to-gas ratio of MgII absorbers]
{On the HI content, dust-to-gas ratio and nature of MgII absorbers}
\author[B. M\'enard  \& D. Chelouche]{Brice M\'enard \& Doron Chelouche\\
\hspace{-.1cm}Canadian Institute for Theoretical Astrophysics}

\begin{document}

\maketitle

\begin{abstract}
  We estimate the mean dust-to-gas ratio of MgII absorbers as a
  function of rest equivalent width $W_0$ and redshift over the range
  $0.5<z<1.4$.  Using the expanded SDSS/HST sample of low-redshift
  Lyman-$\alpha$ absorbers we first show the existence of a $8\sigma$
  correlation between the mean hydrogen column density $\langle
  N_{\rm{HI}}\rangle$ and $W_0$, an indicator of gas velocity
  dispersion. By combining these results with recent dust-reddening
  measurements we show that the mean dust-to-gas ratio of MgII
  absorbers does not appreciably depend on rest equivalent width.
  Assuming that, on average, dust-to-gas ratio is proportional to
  metallicity, we find its redshift evolution to be consistent with
  that of $L^\star$ galaxies from $z=0.5$ to 1.4 and we show that our
  constraints disfavor dwarf galaxies as the origin of such
  absorbers. We discuss other scenarii and favor galactic outflows
  from $\sim L^\star$ galaxies as the origin of the majority of strong
  MgII absorbers.  Finally, we show that, once evolutionary effects 
  are taken into account, the Bohlin et al.  relation
  between $A_V$ and $N_{H}$ is also satisfied by strong MgII systems
  down to lower column densities than those probed in our Galaxy.
\end{abstract}


\section{Introduction}

A successful theory of galaxy formation must not only explain the
properties of the luminous parts of galaxies but it is equally
important to account for the material seen in absorption 
against background sources. 
However,
while the connection between strong absorbers and galaxies was
realized a long time ago
\citep{1969ApJ...156L..63B,1986A&A...155L...8B}, our physical
understanding of absorption selected systems has not yet reached the
maturity of current models addressing the emission properties of
galaxies. In particular, the nature of the structures probed by some
of the strongest absorption features is not yet understood (e.g.
\citealt{2005ARA&A..43..861W}).

Exploring the existence of scaling relations for absorbers may
provide us with useful insight to constrain the physical conditions
and establish theoretical models for these systems and their
associated galaxies.  In this paper we investigate the properties of
the dust-to-gas ratio of MgII absorbers, i.e. the strongest metal line
detectable in optical spectra at $z\lesssim2$ and in some cases a
tracer of Damped Lyman-$\alpha$ absorbers (DLAs)
(e.g. \citealt{2005pgqa.conf...24C}). We first present new
correlations involving gas velocity dispersion, hydrogen and dust
column densities, and show how the knowledge of the dust-to-gas ratio
sheds light on the nature of strong MgII absorber systems.

The dust-to-gas ratio is one of the basic properties of the ISM and
IGM but our knowledge of this quantity is largely based on our own
Galaxy, some of its satellites and a few objects with $z>0$ using for
example multiple images in strongly lensed systems
(\citealt{1997ApJ...477..568Z,2008arXiv0803.1679D}) or absorbers in front of gamma ray
bursts (\citealt{2006MNRAS.372L..38E}). Interestingly, the Milky Way
presents a characteristic value of the extinction per H atom.  Using
hydrogen Lyman-$\alpha$ and $\rm{H_2}$ absorption lines to determine
the total H column densities along star sight lines,
\citet{1978ApJ...224..132B} found that the extinction in the $V$-band follows
\begin{equation}
{\mathrm{A_V}}\simeq 0.53\,
\left(
\frac{N_H} {10^{21}\, {\mathrm{cm}}^{-2}}
\right)
{\mathrm{~mag~~~~~for~~R_V=3.1}}\,,
\end{equation}
with a scatter about the mean of about 30\% and where ${\rm R_V=A_V/E(B-V)}$.
This relation was initially observed over the range of column densities
$10^{20}\lesssim N_H \lesssim 3\times 10^{21}$ cm$^{-2}$ and then
extended and confirmed up to $N_H \sim 5\times 10^{22}$ cm$^{-2}$
where molecular hydrogen plays an important role
\citep{2002ApJ...573..662S}.
The linearity and slope of the above
relation carry information on the complex mechanisms responsible for
the formation and destruction of dust grains. It is interesting to
investigate whether such a relation holds at lower column densities
(where ionization corrections could be important) and/or in different
environments such as larger galactic radii and higher redshifts.

By extrapolating the above relation, column densities with
$N_H\lesssim10^{20}$ cm$^{-2}$ are expected to be associated with
E(B-V) values smaller than a few$\,\times10^{-2}$ magnitude.
Measuring such an effect
along individual star sight lines becomes challenging and a statistical
approach is required.  Such a technique was recently used 
with DLAs by \citet{2008A&A...478..701V} and MgII absorbers
by \citet{York+06}, \citet{Wild06} and \citet{2008MNRAS.385.1053M}.
In particular, the latter authors constrained the reddening induced by
MgII absorbers down to E(B-V)$\sim5\times10^{-3}$ mag 
by analyzing $\sim 7000$ intervening systems.

By combining reddening constraints of MgII absorbers with measurements
of hydrogen column densities, the present analysis will allow us to
obtain constraints on dust properties in new regimes, i.e. at $z\sim
1$ and on large scales ($\sim 50$ kpc) around galaxies.  In particular
we will show that the inferred dust-to-gas ratio of MgII is similar
to that given by the
\cite{1978ApJ...224..132B} relation extrapolated down to lower 
hydrogen column densities.

Throughout the paper we will simply refer the dust-to-neutral gas
ratio as dust-to-gas ratio unless stated otherwise. The amount of dust will be
quantified by either E(B-V) or A$_V$. The MgII rest
equivalent width, $W_0$, denotes the 2796$\,$\AA\ transition.

\section{THE DATA}

The data used in this analysis come from two sources: (i) ``the
expanded SDSS/HST sample of low-redshift Lyman-$\alpha$ absorbers''
compiled by Rao et al. (2006) and (ii) the characterization of the
reddening effects induced by MgII absorbers in the SDSS by
{\citet{2008MNRAS.385.1053M}. We briefly describe them below.\\

The expanded SDSS/HST sample of low-redshift Lyman-$\alpha$ absorbers
consists of 197 systems and represents the largest sample of
UV-detected DLAs ever assembled.  It has been compiled from UV spectra
of QSOs with MgII systems optically identified in the range
$0.11<z<1.65$.  The MgII absorber systems were selected from various
sources in the literature (See Rao. et al 2006 for references).  Most
of the UV data were obtained with observing programs on HST led by
S. Rao as well as archival data.

In Fig. \ref{plot_rao_data} we show the distribution of neutral
hydrogen column density, $N_{HI}$, as a function of MgII rest equivalent
width, with $0.45<W_0<3.3\,$\AA.  The arrows show upper limits on
$N_{HI}$.  They represent about 3\% of the data points and can safely
be neglected in the present analysis.  An important point to emphasize
is the large scatter in $N_{HI}$ (up to three orders of magnitude)
even at a fixed value of $W_0$.  The color-coding of the points refer
to the measured rest equivalent width of FeII at 2600 \AA. 
Empty circles are used for upper limites.

The second set of observational results used in our analysis comes
from the recent dust-reddening analysis done by
{\citet{2008MNRAS.385.1053M}.  Using almost 7,000 MgII absorbers
detected in SDSS quasar spectra these authors have measured the mean
color excess E(B-V) induced by these systems, as a function of rest
equivalent width and redshift. Such constraints will be used to
compute the dust-to-gas ratio of MgII absorbers.

It is useful to recall that the majority of MgII absorption lines with
$W_0\gtrsim1$~\AA\ are saturated \citep{Nestor+05} and that, therefore,
the rest equivalent width of the lines provides us with an estimate of
the gas velocity dispersion.  For a completely saturated line,
$1\,$\AA\ corresponds to $\Delta v\simeq$ 107 km/s. Using
high-resolution spectroscopy the absorption is often seen to originate
from several velocity components implying that this value is only a
lower limit on the total velocity dispersion of the
system. Empirically it has been observed that $\Delta v\simeq 120$ km
s$^{-1}\,$\AA$^{-1}$ (\citealt{2006MNRAS.368..335E}, Fig. 3).  This
value will be used below for physical interpretations.

\begin{figure}
  \includegraphics[width=\hsize]{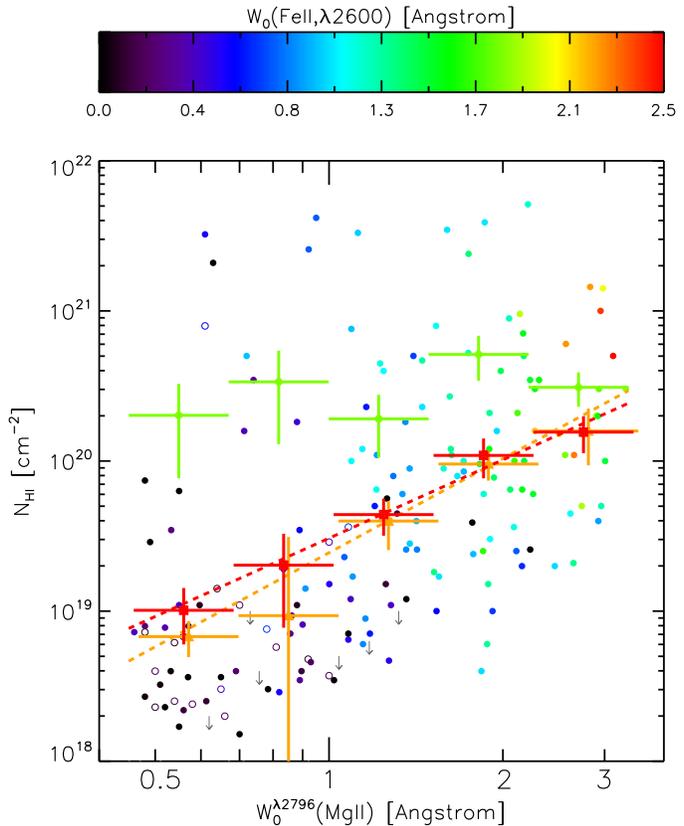}
\caption{Distribution of hydrogen column densities $N_{HI}$ of MgII
  selected absorbers, as a function of rest equivalent width
  $W_0^{2796}$. The data originate from the expanded SDSS/HST sample
  of low-redshift Lyman-$\alpha$ absorbers compiled by Rao et
  al. (2006).Typical errors are about 0.1 dex in ${\rm N_{HI}}$ 
  and about $0.07\;\rm \AA$ in $W_0^{2796}$.
  The arrows show upper limits in ${\rm N_{HI}}$, the color of
  the points refer to the measured rest equivalent width of FeII,
  $W_0^{2600}$. Empty circles denote upper limits of $W_0^{2600}$. The
  green, red and orange points with error bars represent the
  arithmetic mean, geometric mean and median in each bin,
  respectively.}
\label{plot_rao_data}
\end{figure}

\section{The distribution of hydrogen column densities}

The expanded SDSS/HST sample of low-redshift Lyman-$\alpha$ absorbers
was used by \cite{Rao+06} to quantify the amount of neutral hydrogen
probed by MgII absorbers at $z<1.65$ and to estimate the mean hydrogen
density $\Omega_H$.  To do so, these authors computed the mean HI
column density as a function of $W_0(\rm{MgII})$:
\begin{equation}
\big\langle {\rm N_{HI}} \big\rangle(W_0)=
\frac{1}{N}\,\sum_{i=1}^{N}\,N_{HI,i}\,,
\end{equation}
and found $\langle {\rm N_{HI}} \rangle \sim 10^{21}\,{\rm cm}^{-2}$
for the range $W_0>0.6\,$\AA.  In Figure \ref{plot_rao_data} we show
the arithmetic mean of ${\rm N_{HI}}$ in green, using a logarithmic
binning in $W_0$.  We have used 500 bootstrap samples in
order to estimate the errors on the mean in each bin.  We recover the
lack of correlation between mean hydrogen column density reported by
\cite{Rao+06}. For $W_0>0.6\,\rm{\AA}$ we find $\langle {\rm N_{HI}}
\rangle \simeq 3\times 10^{20}\,{\rm cm}^{-2}$.

While the use of the arithmetic mean is needed to quantify
the amount of neutral hydrogen probed by MgII absorbers, which is
important to constrain the evolution of HI through cosmic time, it
ends up being sensitive to a small fraction of the data points, having
the largest $N_H$ values. Indeed, the distribution $P(N_{HI}|W_0)$ is
highly asymmetric and spans about three orders of magnitude in the
present dataset. The value of $\langle N_{HI} \rangle$ is therefore
driven by a small fraction of the data points. Even if it is computed
from a sample of $\sim$200 objects, it might effectively suffer from
small number statistics.
\footnote{The same problem affects estimates of the cosmological
  density of neutral gas, $\Omega_g(z)$, as its estimation involves a
  similar average (Lanzetta et al. 1991):
\begin{eqnarray}
\Omega_g(z)=\frac{H_0}{c}\,\frac{\mu\,m_H}{\rho_{crit}} \frac{\sum_i
  N_{HI,i}}{\Delta X}\,,\nonumber
\end{eqnarray}
  where $\mu$ is the mean molecular weight of the gas, $m_H$ is the
  mass of the hydrogen atom, $\rho_{crit}$ is the current critical
  mass density and $\Delta X$ is the absorption distance path.  This
  \emph{effective} small number statistic might be at the origin of
  the discrepancies in $\Omega_{HI}$ currently debated in the
  literature.  Based on a sample of $\sim20$ MgII selected systems
  with $N_{HI}$ measurements, \cite{2004MNRAS.352.1291P} already
  reported such an effect.}

More important, the arithmetic mean of $N_{HI}$ does not provide
relevant information regarding the majority of MgII absorbers and is
therefore not suited to extract underlying correlations between the
parameters $N_{HI}$ and $W_0$. However, additional information
\emph{can} be extracted and we first consider the geometric mean of
$N_{HI}$:
\begin{eqnarray}
\big\langle {\rm N_{HI}} \big\rangle_g(W_0)&=&
10^{{\langle \log N_{HI} \rangle}}\,.
\end{eqnarray}
We compute it using the same binning in $W_0$ and the errors are again
estimated with bootstrap resampling. The results are shown in Figure
\ref{plot_rao_data} with the red data points. We can now detect a
strong correlation between the hydrogen column density $N_{\rm{HI}}$
and the MgII rest equivalent width $W_0$ and a simple
power-law fit gives
\begin{equation}
\big\langle {\rm N_{HI}} \big\rangle_g(W_0)=
C_g\,\left(W_0\right)^{\alpha_g}
\label{eq_geo}
\end{equation}
where the subscript $g$ denotes a geometric mean,
$C_g=(3.06\pm0.55)\times 10^{19}$ cm$^{-2}$ and
$\alpha_g=1.73\pm0.26$. This correlation holds over one order of
magnitude in $W_0$. Its significance, quantified by that of the slope, is
greater than 6$\sigma$.\\

To demonstrate the robustness of the above result, we repeat the
procedure and estimate the median of $N_{HI}$ as a function of
$W_0$. The results are shown in Figure \ref{plot_rao_data} with orange
points. They show consistency with the results obtained using the
geometric mean. Similarly, a power-law fit gives
\begin{equation}
\mathrm{med}\left[{\rm N_{HI}} \right](W_0)=
C_m\,\left({W_0}\right)^{\alpha_m}
\label{eq_med}
\end{equation}
with $C_m=(2.45\pm0.38)\times 10^{19}$ cm$^{-2}$ and
$\alpha_m=2.08\pm0.24$. The correlation is now detected at 8.7$\sigma$.
Consistent fitting parameters are obtained
using both estimators which show that, even if there is a large
scatter between $N_{HI}$ and $W_0$, there exists an underlying and
well defined correlation between them. Both the geometric mean and the
median allow us to measure a signal coming from the majority of the
systems, and not from a few outliers.  It is interesting to mention
that the correlation weakens as we increase the lower limit for $N_H$,
but it is not severely affected if we decrease the higher limit of
$N_H$ (the opposite statement would apply to the arithmetic mean). Our
results indicate that, typically, $N_{HI}$ is roughly proportional to
the square of $W_0$.\\

Similarly, we can quantify the relations between $N_{HI}$ and FeII
rest equivalent width.  By doing so we should keep in mind that the
sample used in this analysis is MgII-selected. The corresponding
FeII-related correlations are valid in this context only.

We have applied the three estimators introduced previously and
summarize the results in Table \ref{table_summary}. The correlation
between the two parameters is detected at a high significance (8 and
12$\sigma$) and is closer to a simple
proportionality 
\footnote{We also note that a double power-law is a more accurate
  representation of the data in the $(N_{HI},W_0(FeII))$ plane.
  This is certainly an interesting feature to address but is beyond
  the scope of this paper, focusing on MgII absorption.}
. Such a
behavior is expected as the $2600\,$\AA\ FeII absorption line has a
lower oscillator strength that the MgII doublet. It is therefore less
subject to saturation and carries more information about the column
density of the absorbing system.

\begin{table}
\caption{Scaling parameters}
\begin{center}

 \begin{tabular}{lcc}
  \hline
    & Amplitude & Power-law index\\
    & [atom.cm$^{-2}$] & \\
    \hline
    $N_H$ vs W$_0$(MgII)\\
    arithmetic mean &$(2.37\pm0.63)\times10^{20}$ &$0.32\pm0.35$\\
    geometric mean &$(3.06\pm0.55)\times10^{19}$ &$1.73\pm0.26$\\
    median &$(2.45\pm0.38)\times10^{19}$ &$2.08\pm0.24$\\
    \hline
    $N_H$ vs W$_0$(FeII)\\
    arithmetic mean &$(2.49\pm0.44)\times10^{20}$ &$1.54\pm0.12$\\
    geometric mean &$(7.85\pm1.14)\times10^{19}$ &$1.34\pm0.11$\\
    median &$(5.74\pm1.33)\times10^{19}$ &$1.46\pm0.18$\\
    \hline
  \end{tabular}

\label{table_summary}
\end{center}
\end{table}

\section{Dust-to-gas ratio}

The presence of dust associated with MgII absorbers has been reported
by several authors
\citep{mp03,2004ApJ...609..589W,Khare+05,York+06,Wild06}. Recently,
\citet{2008MNRAS.385.1053M} analyzed close to 7,000 strong MgII
absorbers and quantified the amount of reddening as a function of rest
equivalent width and redshift.  In particular, they find a scaling
relation between the amount of reddening and the MgII rest equivalent
width:
\begin{equation}
\langle E(B-V)_{\rm rest} \rangle\,(W_0,z)=
\rm C \times \left(\frac{W_0}{1\rm\AA}\right)^\alpha\,(1+z)^\beta\,,
\label{eq_EBV}
\end{equation}
where $\alpha=1.88\pm0.17$, $\beta=-1.1\pm0.4$ and
$C=(0.60\pm0.07)\times10^{-2}$ mag.  This relation has been
constrained over the range $1<W_0<6\,$\AA.  The similarity of
Eq. \ref{eq_EBV} and Eq. \ref{eq_geo} is striking and suggests that
the average dust-to-gas ratio of MgII selected systems does not
strongly depend on rest equivalent width.  We now quantify this
statement by measuring the dust-to-gas ratio in the range of MgII rest
equivalent widths common to both datasets, i.e. $1<W_0<3.3\,$\AA.

To do so we have rerun the reddening analysis by
\citet{2008MNRAS.385.1053M} selecting only MgII absorbers with
$0.4<z<1.4$ so that the redshift distributions of the two datasets are
similar.  It is important to note that, the median and arithmetic
means give consistent results for the reddening analysis 
\footnote{The geometric mean cannot be computed as 
the signal originates from positive and negative fluctuations.}.
This is probably due
to the fact that high dust column density systems cannot be detected
in the SDSS survey due to the extinction bias and therefore the tail
of the reddening distribution cannot be probed.  Richards et
al. (2003) and \citet{2008MNRAS.385.1053M} showed that quasars
reddened by a color excess $E(B-V)$ greater than about 0.3 magnitude
can hardly be selected.  Fortunately, this effect is not expected to
strongly bias the reddening constraints as \cite{Ellison+04} showed,
using radio-selected quasars, that no more than 20\% of
optically-selected quasars are missed due to extinction effects
associated with strong MgII absorbers.

\subsection{Dependence on $W_0$}
\label{section_w0}

\begin{figure}
\begin{center}
  \includegraphics[width=1.\hsize]{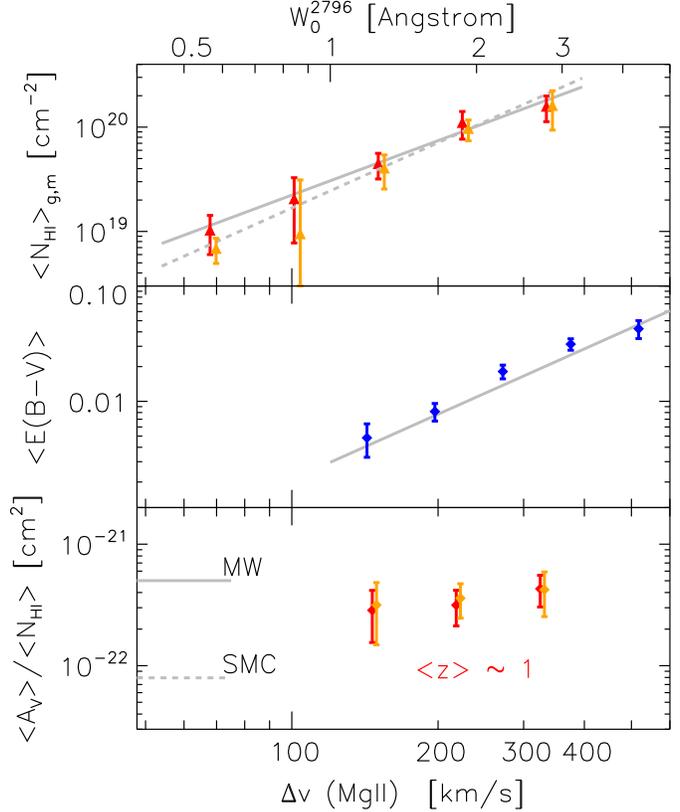}
  \caption{\emph{Upper panel:} geometric mean and median estimates of
  $N_{HI}$ as a function of MgII rest equivalent width (upper x-axis)
  or gas velocity width (lower axis). \emph{Middle panel:} Mean
  reddening induced by MgII absorbers from 
  \citet{2008MNRAS.385.1053M}. 
  \emph{Lower panel:} The dust-to-gas ratio $A_V/N_{HI}$
  which does not significantly depend on the MgII gas velocity width.}
\label{plot_D_G}
\end{center}
\end{figure}
We present our measurements of the dust-to-gas ratio of MgII absorbers
in Fig. \ref{plot_D_G}. The upper panel show the variation of the
geometric mean and median of $N_{HI}$ as a function of MgII rest
equivalent width (upper axis) or $\Delta v(MgII)$, the gas velocity
dispersion (lower axis). The lines show the fitted power-laws
described in Table 1.  The middle panel shows the observed reddening
values $E(B-V)$ for MgII absorbers selected with $0.4<z<1.4$. The
solid line is the fitting formula (Eq. \ref{eq_EBV}) proposed by
\citet{2008MNRAS.385.1053M} and simply evaluted at the mean redshift
of the sample.  To present the results in convenient units we convert
$E(B-V)$ reddening values into visual extinction $A_V$ using a small
Magellanic cloud (SMC) extinction curve, i.e. $R_V=3.1$ as motivated
by \cite{York+06} and \citet{2008MNRAS.385.1053M}.  We use the ratio
between $<A_V>$ and $<N(HI)>$ (directly obtained from the data) as an
estimate of the mean dust-to-gas ratio of MgII absorbers
\footnote{Ideally, the average dust-to-gas ratio $<A_V/N(HI)>$ can be
  estimated by weighting the individual reddening measurements by
  1/N$(HI)$, as done by \cite{2008A&A...478..701V}. In the present
  study, the number of objects for which N$(HI)$ is available is
  unfortunately too small to provide us with a reddening detection.}.
For systems with $1<W_0<3.3\,\mathrm{\AA}$, we find using the geometric mean
or the median of N$(HI)$:
\begin{equation}
  \frac{\left\langle A_{V} \right\rangle}{\left\langle N(\mathrm{HI}) 
    \right\rangle}  = 3.0\pm0.6
  \times10^{-22}~\rm{mag~cm^{2}}
\end{equation}
at a mean redshift close to unity. This value is less than a factor
two lower than that of the Milky Way.  It is interesting to note that
\emph{while the shape of the mean extinction curve of MgII absorbers
  is consistent with that of the SMC}, i.e. does not present the 0.2
$\micron$ bump \citep{York+06,2008MNRAS.385.1053M}, \emph{the mean
  dust-to-gas ratio is substantially higher than that of the SMC}.

Interestingly, the mean dust-to-gas ratio does not appear to be a
strong function of $W_0$. While both the mean hydrogen and dust column
densities vary by more than an order of magnitude over this range, the
mean dust-to-gas ratio is consistent with being constant and is found
to vary by less than a factor $\sim2$ in this interval
for the current sample. In comparison, the dust-to-gas ratios of the
Milky Way and the SMC differ by a factor $\sim8.5$. A power-law fit to
the observed dust-to-gas ratio as a function of MgII rest equivalent
width gives
\begin{equation}
\label{trend_dg}
\frac{\left\langle A_{V} \right\rangle}{\left\langle N(\mathrm{HI}) \right\rangle}
\propto (W_0)^{\gamma}
 \left\{
    \begin{array}{ll}        
 	\gamma_{g}=0.5\pm0.7\\
        \gamma_{m}=0.4\pm0.8\\
    \end{array}
\right.
\end{equation}
A large scatter in the relation between dust-to-gas ratio and MgII
rest equivalent width may exist if we consider individual systems,
however our results show that the \emph{mean} dust-to-gas ratio of
MgII-selected systems is not a strong function of gas velocity
dispersion.  The lack of strong correlation is an interesting
property.  

If we assume that the dust-to-gas ratio is, on average, proportional
to metallicity, the trend reported in Eq. \ref{trend_dg} indicates
that the mean metallicity of MgII absorbers is a weak function
of rest equivalent width.  The linearity between dust-to-gas ratio and
metallicity is expected as dust is formed from metals and the mass of
metals is equal to the product of the metallicity and the gas mass.
This linear correlation has been shown in nearby galaxies
\citep{1990A&A...236..237I,2004A&A...424..465B} or in high redshift
DLAs \citep{2006A&A...454..151V} over several orders of magnitude in
column densities.

Given the existence of a stellar mass-metallicity relation
\citep{2004ApJ...613..898T}, the apparent lack of correlation between the mean dust-to-gas ratio and the velocity dispersion of the gas implies that the mean mass of galaxies giving rise to strong MgII absorption does not play a dominant role in determining $W_0$. In other words, the velocity dispersion of the gas does not reflect the gravitational potential of the system.
By analyzing the correlation between the mean luminosity of MgII
absorbing galaxies and $W_0$, several authors have reported similar
results. 
Using a sample of 58 MgII absorber-galaxy associations, \cite{Steidel+97}
did not find any correlation between galaxy luminosity and absorber rest
equivalent width. Similar results were also
found by \citet{Zibetti+07} who stacked the images of $\sim2500$ SDSS
quasars with strong MgII absorbers and \citet{K+2007} who reported a
lack of correlation between MgII absorption and galaxy morphology from
HST data.\\

The present analysis allows us to probe only a factor $\sim3.3$ in
MgII rest equivalent width, which corresponds to velocity widths in
the range $120 \lesssim \Delta v \lesssim 400$ km s$^{-1}$.  The
number of absorbers per unit rest equivalent width and redshift is
given by Nestor et al. (2005):
\begin{equation}
\partial N/\partial W_0 = \frac{N^*}{W^*}
e^{-\frac{W_0}{W^*}}\,,
\end{equation}
with the maximum likelihood values $W^*=0.702 \pm 0.017$~\AA\ and
$N^*=1.187 \pm 0.052$. The steep exponential decline implies that the
incidence of systems with $W_0\simeq1\,\mathrm{\AA}$ is about 30 times
higher than that of systems with $W_0\simeq3.3\,\mathrm{\AA}$.
  The sharp decrease of dN/d$W_0$ may be due to a transient nature of
  the absorbing gas, such as outflows triggered by star formation.\\

These results may, at first, appear to be in contrast with direct
measurements of metallicities and MgII rest equivalent widths reported
by \citet{2007MNRAS.376..673M}. These authors found $\langle Z \rangle
\simeq (1.69\pm0.20)\,\log(W_0)+cst$ which indicates a substantially
steeper dependence on $W_0$.  However, their sample selection
significantly differs from ours: their study is not based on
  MgII selected absorbers in general, but focuses only on systems
with both detections of hydrogen column density and a metallic
absorption line from a volatile element.  As a result of this
additional selection criterion, the distribution of points in the
$(N_{HI},W_0)$ plane differs from the generic one compiled by Rao et
al. (2006) and presented in Fig. \ref{plot_rao_data}.  For example, at
$W_0<1\,\mathrm{\AA}$ the region of the plane with
$N_{HI}<10^{21}\,\mathrm{cm}^{-2}$ is almost unpopulated in their case
while it is where most of the points lie in the Rao et al. sample.
Hence, the metallicity trend reported by Murphy et al.  only applies
to a specific sub-population of MgII absorbers.

We postulate that, similarly to the shallow dependence found between
dust-to-gas ratio and $W_0$, a weak correlation is expected between
metallicity and $W_0$ providing that absorbers are selected only on
their MgII rest equivalent width.  Such a measurement can be performed
by using composite spectra
(e.g. \citealt{2003ApJ...595L...5N,2005astro.ph..6701T,York+06}) which
allows one to detect weaker absorption lines and estimate mean
metallicities.

\begin{figure}
\begin{center}
  \includegraphics[width=1.1\hsize]{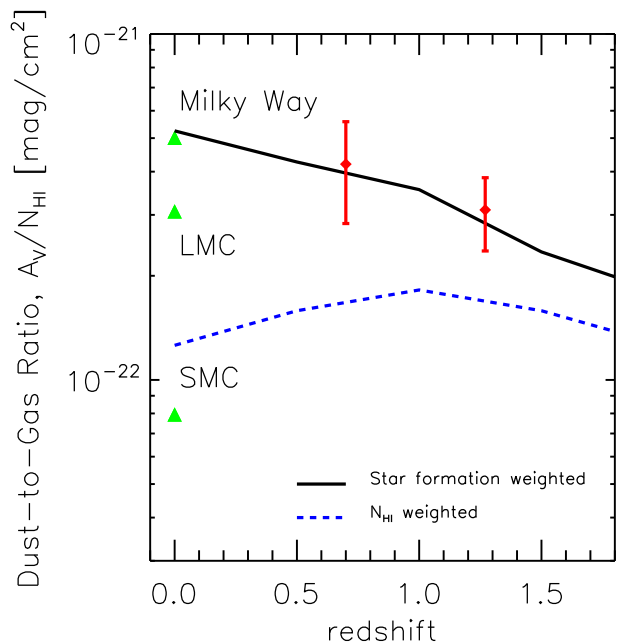}
  \caption{The mean dust-to-gas ratio of strong MgII absorbers as a
  function of redshift. For comparison we show an estimate of the
  expected dust-to-gas ratio of $\sim L^\star$ galaxies from
  Dav\'e \& Oppenheimer (2007)  
  as well as the values of the Milky Way, the LMC and the SMC \citep{2003ApJ...594..279G}.}
\label{plot_redshift_evolution}
\end{center}
\end{figure}

\subsection{Redshift evolution and link to $L^\star$ galaxies}

Having shown that the dust-to-gas ratio of MgII absorbers does not
strongly depend on rest equivalent width (over the range
$1<W_0<3.3\,\mathrm{\AA}$), we now investigate its evolution as a
function of redshift. Following the above procedure, we have divided
both datasets into two redshift bins with $\langle z_1 \rangle \simeq
0.7$ and $\langle z_2 \rangle \simeq 1.2$ and measured the mean
$A_V/N_{HI}(z)$ as defined above. We present the results in
Fig. \ref{plot_redshift_evolution} and the numerical values are given
in table 2.  As can be seen, the observed
dust-to-gas ratio of strong MgII systems is, on average, similar to
that of the Milky Way and significantly higher than that of the SMC,
even at the highest redshift we can probe.  The data also suggest a 
trend of decreasing dust-to-gas ratio with increasing redshift.

\begin{table}
\caption{Dust-to-gas ratio of MgII absorbers with $1<W_0<3\,{\rm \AA}$}
\begin{center}

 \begin{tabular}{lc}
  \hline
  \hline
    & ${\rm A_V/N_{HI}~[10^{-22}\;mag/cm^2]}$ \\
    \hline
    $\langle z \rangle = 0.7$ & $4.20 \pm 0.14$\\    
    $\langle z \rangle = 1.2$ & $3.10 \pm 0.74$\\
    \hline
    \hline
  \end{tabular}

\label{table_summary}
\end{center}
\end{table}

As done previously, assuming that the dust-to-gas ratio is, on
average, proportional to the metallicity (see section
\ref{section_w0}), we can attempt to compare the observed trend to
available models. To do so, we first consider that a solar metallicity
corresponds to a Milky Way dust-to-gas ratio
{\citep{1978ApJ...224..132B,1990A&A...236..237I}:
\begin{equation}
\left \langle
\log \left[{\rm \frac{A_V}{N_{HI}}}(z) \right] 
\right \rangle
\simeq
\log{\rm \left( \frac{A_V}{N_{HI}} \right)_{MW}}\,+
{\rm \left\langle \left[ Z/H \right] \right\rangle }(z)\,.
\label{eq_dg}
\end{equation}
The above estimate strongly depends on the averaging procedure used to
define the mean metallicity.  Modelled evolutions of ${\rm
  \left\langle \left[ Z/H \right] \right\rangle}(z)$ have been
explored by \cite{2007MNRAS.374..427D}. 
These authors investigated the cosmic metal budget 
in various phases of baryons estimated from cosmological 
hydrodynamic simulations which included constrained 
models for enriched galactic outflows.
The solid line in Fig. \ref{plot_redshift_evolution} shows their estimation of
the variation of mean metallicity as a function of redshift  for a
star-formation weighted estimator, expected to be a representative
value for galaxies selected in emission.  We can see that, over the
entire redshift range available, the dust-to-gas ratio of
MgII-selected systems is in agreement with the model of the
metallicity evolution using a star formation-weighted estimator
 (representative of $L^\star$ galaxies) but not with the N(HI)-weighted one 
(corresponding to DLAs). Even at the highest redshifts probed by our dataset, the
mean dust-to-gas ratio of MgII absorbers is significantly higher than
that of the SMC at $z=0$.  For comparison, we also show the expected
evolution of an $N_{HI}$-weighted estimator of ${\rm \left\langle
  \left[ Z/H \right] \right\rangle}(z)$ with the dashed line. As can
be seen, such a trend is not in good agreement with the data points.

These results show that the mean dust-to-gas ratio of strong MgII
absorbers is consistent with that of $L^\star$ galaxies but not with
that of substantially smaller systems, such as the SMC. Given
  that metallicity is expected to decrease with increasing redshift,
  an LMC-type dust-to-gas ratio is also disfavored.
It suggests that, on average,
strong MgII systems are associated with $\sim L^\star$ galaxies (which
does not prevent the existence of a large scatter around this relation
if individual systems are considered).

Associations between strong MgII absorbers and $\sim L^\star$ galaxies
have already been made observationally.  As mentioned above,
\cite{Steidel+97} and \citet{Zibetti+07} showed that the mean
luminosity of MgII absorbing galaxies is about
$0.8\,L^\star$. However, while such associations clearly gathered
information on the link between strong MgII absorbers and $\sim
L^\star$ galaxies, they have not been able to provide strong
constraints on the nature of these systems.  Below we show how the
knowledge of the mean dust-to-gas ratio may shed light on the origin
of these systems.

\subsection{Nature of the absorbing gas}

The origin of MgII absorbers has been a matter of debate since their
discovery.  Various scenarii have been proposed: infalling material
\citep{1996ApJ...469..589M}, outflows \citep{2001ApJ...562..641B},
orbiting dwarf galaxies \citep{1986ApJ...311..610Y}, etc.
Certain studies have proposed
a relation between strong MgII absorbers and star formation: from the
observed redshift distribution of dN/d$z$
\citep[e.g.][]{1997A&A...328..499G,2006ApJ...639..766P} or from the
existence of a correlation between MgII rest equivalent width and star
formation rates estimated from broad band colors \cite{Zibetti+07}.
While some indication of the outflow phenomenon exists in certain cases, two
different origins remain: outflows can equally well originate from the
bright galaxy usually found at $\sim50-100$ kpc from the absorber or
from the satellite galaxies orbiting the parent dark matter halo of
the system, usually too faint to be observed and/or too close to the background quasar.

Interestingly, our result allows us to strengthen the outflow hypothesis. 
Considering that
\begin{enumerate}
\item the mean impact parameter of MgII absorbers is about 50 kpc from
  a $L\sim L^\star$ galaxy \citep{Zibetti+07},
\item the mean dust-to-gas ratio of these systems is consistent with
  that of $L^\star$ as a function of redshift,
\end{enumerate}
this suggests that the gas is originating from the neighboring $\sim
L^\star$ galaxy itself. In such a context, outflows appear to be the
simplest explanation to simultaneously explain these two constraints.
Dwarf galaxies alone do not satisfy the second property. Our result
therefore indicates that the bulk of MgII absorbers with $W_0>1\,$\AA\
and $0.5\lesssim z \lesssim 1.5$ may trace outflowing gas from
galaxies.
Another scenario would involve low-metallicity gas outflowing or being
stripped from dwarf galaxies and experiencing a higher dust-to-neutral
gas ratio due to ionization effects. Such a model is however less
attractive as it requires ionization effects to coincidentally
increase the dust-to-neutral gas ratio to the value of that of a
$L^\star$ galaxy.\\

In some cases, observations of individual galaxies have already
pointed out to a link between strong MgII absorbers and outflows: 500
km/s-blueshifted MgII absorption has been observed in the spectra of
10 out of 14 massive post-starburst galaxies
\citep{2007ApJ...663L..77T}. Recently, using integral-field
spectroscopy, \cite{2007ApJ...669L...5B} probed H-$\alpha$ emission
within $30$ kpc of MgII absorbers with $W_0>2\,$\AA\ at $z\simeq0.9$,
and found a 2-$\sigma$ indication of correlation between MgII rest
equivalent width and H-$\alpha$ flux based on 14 systems.  Using
high-resolution spectroscopy of a few strong MgII systems,
\cite{2001ApJ...562..641B} showed that the shape of certain absorption
lines is consistent with those expected from galactic winds (but could
also be due to merging galaxies).   While the above analyses
  dealt with individual systems, the statistical results presented in
  our analysis \emph{suggest} that gas traced by MgII absorption, with
  a significant range of MgII rest equivalent widths and over a
  substantial fraction of cosmic time, may often be associated with an
  outflow phenomenon.

\subsubsection*{The diversity of MgII-selected systems}

\begin{figure*}
  \includegraphics[width=.8\hsize]{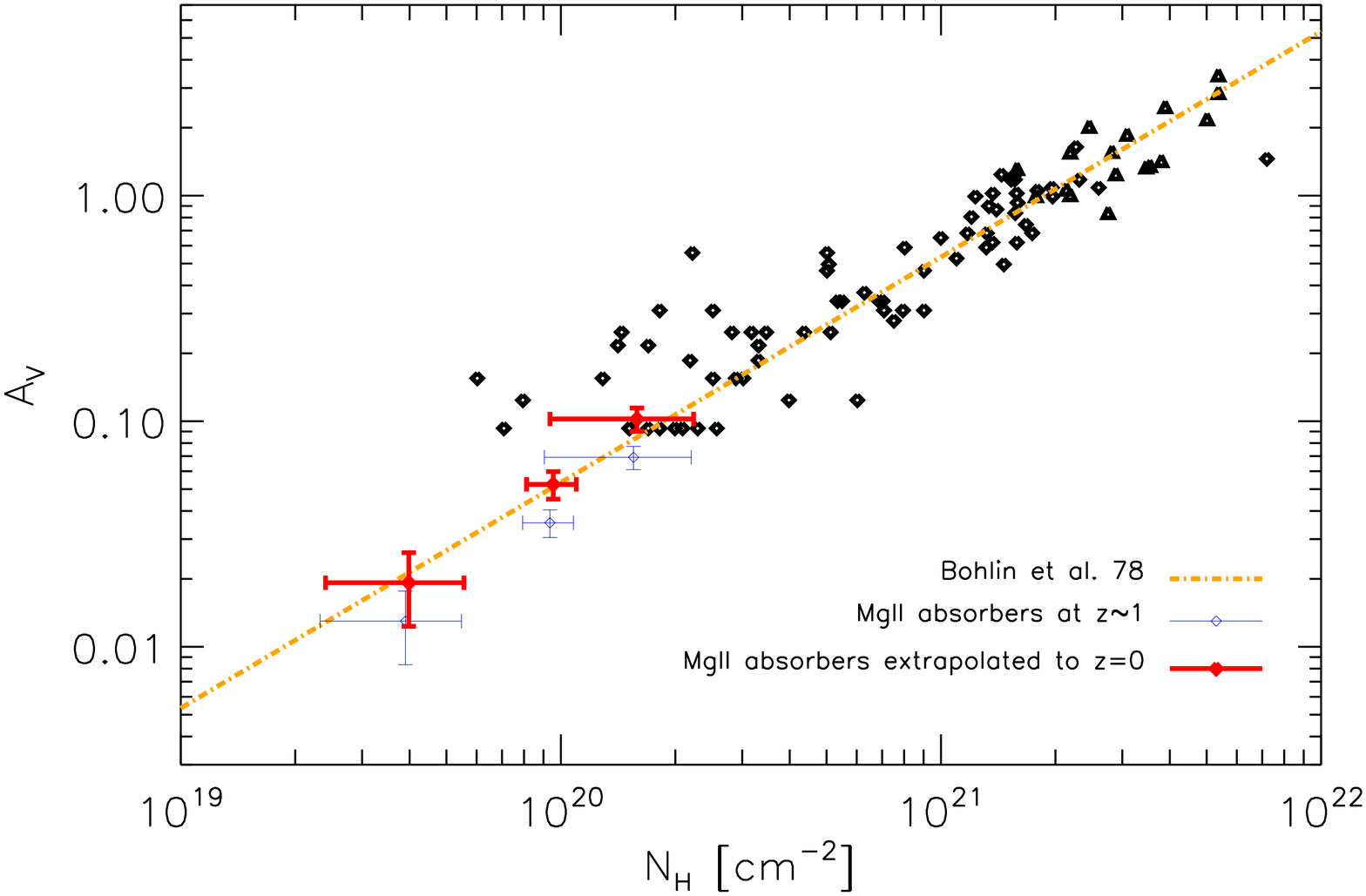}
\caption{Relation between visual extinction $A_V$ and neutral hydrogen
  column density, $N(HI)+2\,N(H_2)$. Data points with diamonds were
  taken from \citet{1986ApJ...301..355J} and with triangles from
  \citet{2002ApJ...573..662S}.  The line shows the best fit obtained
  from lines of sight within our Galaxy, by
  \citet{1978ApJ...224..132B}.  The data points with error bars show
  the median $N_{HI}$ and $A_V$ for MgII absorbers with
  $1.<W_0<1.48,~1.48<W_0<2.21,~2.21<W_0<3.3$ by increasing column
  densities. Blue points show direct measurements (with $\langle z
  \rangle \sim 1$) and red points show the extrapolated values to
  $z=0$ taking into account metallicity evolution.}
\label{plot_Bohlin}
\end{figure*}

MgII absorption arises in gas spanning several decades of neutral
hydrogen column density and probes a wide range of environments.
Locally MgII absorbers are known to trace diverse structures such as
the disk of our Galaxy \citep{1996ApJ...464..141B}, in high velocity
clouds \citep{1995ApJ...448..662B,2000ApJS..129..563S} or the Large
Magellanic cloud \citep{1999ApJ...512..636W}.

The use of the geometric mean or median hydrogen column density and
dust-to-gas ratio has allowed us to estimate number-count weighted
quantities rather than $N_{HI}$-weighted of a MgII-selected
population.  Our results aim at representing ``typical'' systems
but cannot be extended to all objects.
The distribution of $N_H$ as a function of $W_0$ seems to indicate the
existence of two different populations of objects: most of the points
lie around the dashed lines obtained using the geometric mean or
median estimates of $\langle N_H \rangle$ however, others depart by
two to three orders of magnitude in $N_{HI}$ from these relations and
reach the DLA regime.  Such systems are likely to have a different
nature.  First, we can observe that the absorbers with the largest
hydrogen column density do not display the strongest FeII rest
equivalent width (which better correlates with metal column density
than MgII).  Furthermore, their implied dust-to-gas ratio is
necessarily low otherwise the background quasar would be heavily
extincted. We thus postulate that these systems correspond, on
average, to objects substantially less metal-rich such as dwarf
galaxies.
This statement is in line with the mean dust-to-gas ratio and
metallicity estimates of DLAs by \citet{2008A&A...478..701V}.
Using about 250 DLAs with $2.2<z<3.5$ detected in SDSS quasar spectra,
these authors found ${\langle A_{V}/N(\mathrm{HI})\rangle}
  \simeq 2$ to $4\times 10^{-23}$ mag cm$^{2}$, i.e. a dust-to-gas
ratio an order of magnitude lower than the mean values of MgII
absorbers at $z\sim 1$.  As metallicity evolution is not expected to
be as large between redshifts 1 and 2 (see
\citealt{2007MNRAS.374..427D} or the models summarized in
\citealt{2006MNRAS.372..369P}), it indicates that the data points
lying in the upper part of Fig. \ref{plot_rao_data} have an average
dust-to-gas ratio significantly lower than those following the median
track and are likely to trace dwarf-like galaxies. As such, the
absorbers deviating from the median track and populating the upper
part of Fig. \ref{plot_rao_data} should correspond to lines-of sight
intercepting galaxies less metal-rich and probably less luminous than
$L^\star$.  It will be of interest to test this prediction with
available or upcoming datasets. A similar conclusion was reached by
Khare et al. (2007) based on the implied relation between $A_V$ and
$N(ZnII)$ for MgII selected systems from \citet{York+06}. Their
analysis assumed an SMC dust-to-gas ratio. It is worthwile to
re-explore their statement for a higher dust-to-gas ratio as suggested
by the present analysis.

\section{The $\mathrm{A_V}$-$\mathrm{N_H}$ relation}

As mentioned in the introduction, the \cite{1978ApJ...224..132B}
relation, $A_V/N_H\simeq 5.3\times 10^{-22}$ mag cm$^2$ has been
established from measurements with E(B-V)$>0.1$ mag. As our
statistical approach allows us to be sensitive to reddening values
lower by an order of magnitude, it is interesting to investigate the
validity of the such a relation for (i) lower column densities and
(ii) in extragalactic environments. In Fig. \ref{plot_Bohlin} we show
measurements of $A_V$ and $N_H$ for lines-of-sight within our Galaxy
from \cite{1978ApJ...224..132B} and \cite{2002ApJ...573..662S}.  The
blue data points show our estimates of $\langle A_V \rangle$ and the
median $N_H$, for three bins of MgII rest equivalent width. The red
points show the same quantity but scaled to $z=0$ using the star
formation weighted metallicity evolution given by
\cite{2007MNRAS.374..427D} and introduced in the previous section.
Remarkably, we find that the Bohlin et al. relation between $A_V$ and
$N_H$ holds down to substantially lower hydrogen and dust column
densities, regardless of ionization corrections.  
Once evolutionary effects are taken into account,
the hydrogen and dust content of strong metal absorbers at $z\sim1$
appear to be in reasonable agreement with this relation.

\section{Summary and outlook}

The expanded SDSS/HST sample of low-redshift Lyman-$\alpha$ absorbers
compiled by Rao et al. (2006) has provided us with the distribution of
hydrogen column densities of strong MgII absorbers.  Using this
dataset, we have shown that:
\begin{itemize}
\item Due to the nature of the distribution (spanning 3 orders of
  magnitude in $N_{HI}$ estimates at a fixed $W_0$), the arithmetic
  mean $\langle N_{HI} \rangle$ is effectively sensitive to only a few
  percents of the systems in a MgII selected sample.  Even if $\langle
  N_{HI} \rangle$ or $\Omega_{HI}$ is computed from a sample of ~200
  objects, it effectively suffers from small number statistics. This
  fact might be at the origin of the discrepancies in $\Omega_{HI}$
  currently debated in the literature.
  
\item Interestingly, despite the large scatter in hydrogen column
  density at a fixed $W_0$, a well defined relation
  exists between these two quantities: both the median and the
  geometric mean of $N_{HI}$ show, at the 6-8$\sigma$ level, that
  $\langle N_{HI} \rangle_{g,m} \propto W_0^\alpha$ with $\alpha \sim
  1.8$ over a decade in MgII rest equivalent width.
\end{itemize}

By combining these hydrogen column density estimations with recent
reddening measurements from \citet{2008MNRAS.385.1053M}, we have shown
that:
\begin{itemize}
\item while the shape of the mean extinction curve of
  $z\sim1$ MgII absorbers is consistent with that of the SMC, the inferred
  dust-to-gas ratio is substantially higher than that of the SMC:
${\left\langle A_{V} \right\rangle}/{\left\langle N(\mathrm{HI}) \right\rangle}
= 3.0\pm0.6  \times10^{-22}~\rm{mag~cm^{2}}$.
\item This dust-to-gas ratio does not strongly depend on
  $W_0$.  It varies by less than a factor two for
  $1<W_0<3.3\,\mathrm{\AA}$\ and suggests that systems in this range
  have a similar origin.  Such a property constrasts with the sharp
  decrease of dN/d$W_0$ which varies by a factor $\sim30$ over the
  same interval. This may be due to a transient nature of the
  absorbing gas, such as outflows triggered by star formation.
\item 
In addition, assuming proportionaly between dust-to-gas ratio and
metallicity, we have shown that the redshift evolution of the
dust-to-gas ratio of MgII systems is in agreement with a star
formation weighted metallicity estimate (from
\citealt{2007MNRAS.374..427D}). Our results therefore
confirm the connection between the bulk of strong MgII absorbers and
$L^\star$ galaxies from dust-to-gas ratio considerations only.
\end{itemize}

Interestingly, given that (i) the mean dust-to-gas ratio of
  these systems does not favor LMC/SMC values and is consistent with
  that of $L^\star$ as a function of redshift, and (ii) the mean
  impact parameter of MgII absorbers is about 50 kpc from a $\sim
  L^\star$ galaxy \citep{Zibetti+07} strongly suggest that the
  absorbing gas originates from the nearby $\sim L^\star$ galaxy.
  Outflowing gas appears as the most simple explanation
  regarding the nature of the majority of strong MgII absorbers.
An alternative scenario would involve low-metallicity gas originating
from dwarf galaxies and experiencing a higher dust-to-neutral gas
ratio due to ionization effects.  Such a model is however less
attractive as it requires ionization effects to coincidentally
increase the dust-to-neutral gas ratio to the value of that of a
$L^\star$ galaxy.\\

Our results made use of geometric mean and/or median estimates.  They
are aimed at representing the bulk of MgII absorbers and are not
sensitive to outliers in the distributions of hydrogen column
densities.  We have shown that the systems populating the upper part
of Fig \ref{plot_rao_data} and corresponding to DLAs are necessarily
less dusty and likely less metal-rich than the bulk of MgII systems
indicated by the median track.  This illustrates that, even if an
overlap exists between the two populations, the majority of
DLA-selected or MgII-selected systems do not probe the same type of
structures.

It will be of particular interest to repeat such an analysis
for metallicity estimates of absorber systems. In addition, accessing several transistions would allow us to better constrain the ionization fraction.

Finally, we have shown that our measurements allow us to probe the
relation between $A_V$ and $N_{H}$, i.e. between dust and hydrogen
column densities, in regimes that were previously unexplored.  In
particular, we have shown that the \cite{1978ApJ...224..132B} relation
measured from lines-of-sight in our Galaxy is also satisfied by MgII
absorbers at $z\sim 1$, for column densities and reddening values
significantly lower than previously probed.  As it relates the amount
of dust to the \emph{neutral} hydrogen column density, this statement
is valid regardless of the level of ionization.\\

If the bulk of MgII absorbers are tracers of outflowing gas, they may
provide us with an interesting view of delayed star formation around
galaxies. Assuming a wind velocity of about 300 km/s, i.e. about 300
kpc/Gyr, the presence of a strong MgII absorber is then typically
related to a burst of star formation which occured about 150 Myr ago.
The observed properties of MgII absorption lines (incidence, redshift
evolution, velocity width, number of components) can provide us with
observational constraints on star formation in the Universe, up to
high redshifts where emission studies are more difficult.

\section*{Acknowledgements}
We thank Giovanni Vladilo, Jacqueline Bergeron, Daniel Nestor and
Sandhya Rao for useful comments on the manuscript.

\end{document}